\documentclass[aip,amsmath,amssymb,reprint]{revtex4-1}

\usepackage [latin1]{inputenc}

\usepackage{graphicx}
\usepackage{dcolumn}
\usepackage{bm}

\usepackage{times,xcolor}
\usepackage[right=0.72in,left=0.72in,top=0.72in,bottom=0.45in]{geometry}

\usepackage[colorlinks,urlcolor=black,citecolor=magenta,linkcolor=blue]{hyperref}

\usepackage{comment}
\usepackage{cleveref}
\usepackage{physics}

\newcommand{\fs}[1]{\textcolor{black}{#1}}

\begin{document}

\title{A low-impedance radio-frequency circuit for fast spin manipulations in cold alkali atoms}
\author{F. Scazza}
\email[Author  to  whom  correspondence  should  be  addressed. E-mail: ] {francesco.scazza@units.it}
\affiliation{Istituto Nazionale di Ottica (CNR-INO), 50019 Sesto Fiorentino, Italy}
\affiliation{European Laboratory for Nonlinear Spectroscopy (LENS), 50019 Sesto Fiorentino, Italy}
\affiliation{Dipartimento di Fisica, Universit\`a degli Studi di Trieste, 34127 Trieste, Italy}
\author{G. Del Pace}
\affiliation{Istituto Nazionale di Ottica (CNR-INO), 50019 Sesto Fiorentino, Italy}
\affiliation{European Laboratory for Nonlinear Spectroscopy (LENS), 50019 Sesto Fiorentino, Italy}
\affiliation{Dipartimento di Fisica e Astronomia, \mbox{Universit\`a degli Studi di Firenze, 50019 Sesto Fiorentino, Italy}}
\author{L. Pieri}
\affiliation{Radioteknos, 50142 Firenze, Italy}
\author{R. Concas}
\affiliation{Istituto Italiano di Ricerca Metrologica (INRiM), 10135 Torino, Italy}
\affiliation{European Laboratory for Nonlinear Spectroscopy (LENS), 50019 Sesto Fiorentino, Italy}
\author{W. J. Kwon}
\affiliation{Istituto Nazionale di Ottica (CNR-INO), 50019 Sesto Fiorentino, Italy}
\affiliation{European Laboratory for Nonlinear Spectroscopy (LENS), 50019 Sesto Fiorentino, Italy}
\affiliation{Department of Physics, Ulsan National Institute of Science and Technology (UNIST), 44919 Ulsan, Republic of Korea}
\author{G. Roati}
\affiliation{Istituto Nazionale di Ottica (CNR-INO), 50019 Sesto Fiorentino, Italy}
\affiliation{European Laboratory for Nonlinear Spectroscopy (LENS), 50019 Sesto Fiorentino, Italy}


\begin{abstract}
We design and implement a low-impedance, high-current radio-frequency (RF) circuit, enabling fast coherent coupling between magnetic levels in cold alkali atomic samples. It is based on a compact shape-optimized coil that maximizes the RF field coupling with the atomic magnetic dipole, and on 
coaxial transmission-line transformers that step up the field-generating current flowing in the coil by a factor $\sim\,4$ to about 7.5\,A for 100\,W of RF driving. This allows to obtain a RF coupling field of about $0.035\,\text{G}/\sqrt{\text{W}}$ at the atomic sample location. 
The system is robust and versatile, as it generates a large RF field without compromising on the available optical access, and its central resonant frequency can be adjusted \textit{in situ}. Our approach provides a cost-effective, reliable solution, featuring a negligible level of interference with surrounding electronic equipment thanks to its symmetric layout. We test the circuit performance using a maximum RF power of 80\,W at a frequency around $82\,$MHz, which corresponds to a measured Rabi 
frequency $\Omega_R/2\pi \simeq 18.5\,$kHz, i.e.~a $\pi$-pulse duration of about $27\,\mu$s, between two of the lowest states of ${}^6$Li at an offset magnetic field of $770$\,G. Our solution can be readily adapted to other atomic species and vacuum chamber designs, in view of an increasing modularity of cold atom experiments.
\end{abstract}

\maketitle


\section{\label{sec:intro} Overview}

Over the past two decades, the ability to coherently manipulate the internal state of atoms and molecules has proven an essential tool in the field of precision measurements \cite{Inguscio2013}, %
as well as for the realization of ultracold synthetic matter and atomic quantum simulators of important many-body problems \cite{Ketterle2008,Varenna2016}.
In particular, a key experimental probe of many-body physics in ultracold atomic systems is radio-frequency (RF) spectroscopy \cite{Torma2016,Ketterle2008,ValeZwierlein}, by which an applied RF pulse is used to transfer atoms from one hyperfine state to another unoccupied state. Addressing the RF transitions that connect hyperfine sublevels with different interaction properties allows for accessing the spectral properties of elementary excitations in Fermi \cite{Chin2004, Stewart2008, Schirotzek2009, Kohstall2012, Koschorreck2012, Scazza2017} and Bose gases \cite{Hu2016, Jorgensen2016, Yan2020, Etrych2025}, as well as fundamental thermodynamic quantities 
\cite{Sagi2012, Mukherjee2019, Fletcher2017}. 
In this context, a significant demand is the implementation of strong coherent RF drives, which are especially important to precisely trigger and probe non-equilibrium many-body dynamics governed by short-range interactions \cite{Bardon2015, Cetina2016, Fletcher2017, Amico2018, Skou2021, Vivanco2025, Etrych2025}, tuned through Feshbach resonances upon varying a static magnetic field applied to the sample. Additionally, strong RF driving notably allows for performing efficient evaporation of atomic quantum gases in magnetic traps~\cite{Ketterle1996} and for dynamically shaping the trapping potentials~\cite{Hofferberth2006,Perrin2017}. It is also becoming increasingly important for realizing fast qubit rotations in atom-based quantum information processing applications~\cite{Braverman2019,Ma2022}.

In quantum degenerate Fermi gases, a natural time scale $\tau_F = \hbar/\varepsilon_F$ is set by the Fermi energy $\varepsilon_F$ of the system, where $\hbar$ is the reduced Planck constant $h/2\pi$. The dynamics at short times $t \sim \tau_F$ are expected to be universal, revealing the emergence of elementary excitations in real time \cite{Cetina2016,Skou2021,Etrych2025}, and serving as a valuable benchmark for many-body theories \cite{FermiPolaronRPP,Adlong2020,Scazza2022}. In order to experimentally investigate such \textit{ultrafast} dynamics in ultracold Fermi gases, it is necessary to control the spin state of the atoms over time scales on the order of $\tau_F$, whose typical values range between $5\,\mu$s and $50\,\mu$s depending on the experimental sample density and atomic mass. 
The experimental time resolution in controlling the atomic spin state is set by the Rabi frequency $\Omega_R$, characterizing the strength of the coupling between an external electro-magnetic (EM) field and the dipole moment associated with the atomic spin. For a given magnetic dipole moment, the only way to increase $\Omega_R$ is to increase the amplitude of the driving field experienced by the atoms. At the same time, it is important that the circuit generating the RF field does not perturb the sensitive electronics within the surrounding apparatus. This makes the realization of quiet and efficient RF systems a priority in cold atom laboratories. Moreover, as experimental setups become more complex, including large-aperture imaging systems and multiple laser beams, critical space constraints call for unobtrusive and versatile RF circuit designs.

Here we present the design and implementation of a flexible, compact and low-cost high-current circuit for efficiently coupling a large RF magnetic field to the spin of alkali atoms, whose transitions within the ground-state manifold \fs{typically lie in the 10 -- 200\,MHz range, allowing for spin manipulations with $\Omega_R \sim \varepsilon_F/\hbar$.} Significant complexity results from the fact that atomic samples are trapped inside an often metallic, ultra-high vacuum 
chamber by laser light potentials, and are subject to a strong static magnetic field $B_0$ produced by Feshbach electromagnets. Fulfilling all design requirements, the realized system shows an excellent electronic performance, in agreement with simulations. 
It has been tested in a typical application case by driving the transition between two ground-state magnetic sublevels of lithium atoms, characterized by a frequency $\nu_0\simeq82$\,MHz. Moreover, it features a very low level of interference with nearby laboratory electronic equipment, as demonstrated by observed near-unity $\pi$-pulse efficiencies even at 100\,W-level RF driving and an estimated static B-field fractional stability of about $10^{-5}$ for $B_0\simeq770$\,G over several minutes of low-power RF driving with a Rabi frequency~$\sim 400 \,$Hz.

The paper is organized as follows: 
in Section II, after recalling the basic concepts behind the coupling between an atomic spin and a classical RF field, we describe our approach to the RF circuit design and its main advantages with respect to more established solutions, and we present the realization of the different components of the RF circuit, namely the coil, the matching network and the transmission-line transformers; in Section III, we report on the characterization of the system performance with an ideal Fermi gas of lithium atoms; in the concluding Section, we summarize and provide some outlook.


\begin{figure*}[!t]
  \begin{center}
    {\includegraphics[width=\textwidth, angle=0]{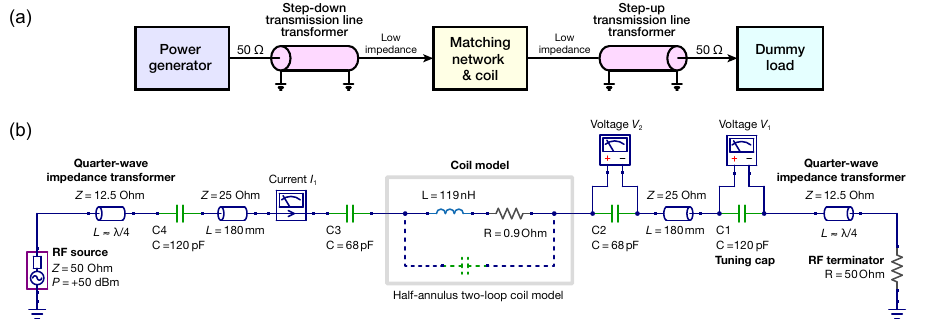} \vspace*{-5pt}}
    \caption{\label{fig:schblkconc} Circuit concept and design. (a) A block diagram displays the five cascaded components composing the RF circuit: the generator, the step-down transformer, the matching network that compensates the coil impedance, the step-up transformer, and the 50$\,\Omega$ dummy load. (b) Full layout of the RF circuit. 
    The step-down and step-up transformers are implemented by quarter-wave 12.5$\,\Omega$ coaxial transmission lines. They respectively decrease and increase the impedance of their ``secondary''  sides by a factor 16.
    The coil is modeled as a LR series, based on impedance measurements of the actual wire loop (see Fig.~\ref{fig:coil}) inserted in a mock-up of the vacuum chamber. The capacitors compose the matching network together with the two 180\,mm-long transmission lines bridging the tuning board and the coil (see also Fig.~\ref{fig:Realiz}). The capacitance values shown here apply to a central frequency around 83\,MHz. \vspace*{-10pt}}
  \end{center}
\end{figure*}


\section{\label{sec:circuit} The circuit}
\subsection{Essentials of atom spin interactions with RF fields}

Energy shifts between hyperfine levels of an alkali atom depend on the applied static magnetic field, which sets the quantization direction $z$, and within the $J=1/2$ ground-state manifold they can be precisely computed through the Breit-Rabi formula\cite{SteckBook,Bransden2003}. Since transitions between hyperfine states or Zeeman sublevels are electric-dipole forbidden, they have an extremely narrow natural linewidth, notably exploited for atomic clocks. External magnetic fields interact with the atom only through the magnetic dipolar interaction\cite{SteckBook,Bransden2003}. In the case of ground-state Zeeman sublevels, the energy splittings for typical $B_0$ values correspond to RF frequencies $\nu_0$ in the so-called very high frequency (VHF) range, roughly between 10 and 100\,MHz, making RF magnetic fields the simplest method to couple different levels with one another. 

Let us consider both static and RF magnetic fields that are spatially homogeneous over the scale of the atomic sample, and that the RF field is linearly polarized. We may define the static offset field as $\mathbf{B}_0 = B_0\,\mathbf{e}_z$, and focus on a pair of adjacent magnetic sublevels, split by an energy $\hbar\omega_0$ associated with $B_0$. These realize an effective atomic spin-1/2 system, as they coincide with the eigenstates $\ket{m}$ of the spin projection $S_z$ along the quantization axis
\footnote{While the considered spin $\mathbf{S}$ typically maps to the total angular momentum $\mathbf{F}$ of the hyperfine atomic state, in the limit of large magnetic fields $B_0$ it approaches the nuclear spin $\mathbf{I}$, which decouples from the electronic shell\cite{Bransden2003}.}. 
The interaction of an atomic spin $\mathbf{S}$ with a RF magnetic field $\mathbf{B}_\mathrm{RF} (t) = B_\mathrm{RF}  \cos(\omega t + \varphi)\,\hat{\epsilon}$ is given by 
$V_\mathrm{RF} = g \mu_B B_\mathrm{RF}\cos(\omega t + \varphi)\,\mathbf{S} \cdot \hat{\epsilon}$, 
where $\mu_B$ is the Bohr magneton, $g$ is the relevant gyromagnetic factor\cite{SteckBook}, $\omega$ and $\varphi$ are the RF B-field angular frequency and phase offset, and $\hat{\epsilon}$ is the field polarization unit vector. 
In order to couple the two $\ket{m}$-levels with one another, the RF field polarization must have non-zero projection in the $x-y$ plane, and its frequency must be close to resonance\cite{SteckBook,Bransden2003}, i.e.~$\omega \simeq 2\pi\nu_0$. 
This is the principle behind the phenomenon of nuclear magnetic resonance (NMR). Taking $\hat{\epsilon} = \mathbf{e}_y$ to maximize the coupling $V_\mathrm{RF}$, without loss of generality we can define the Rabi frequency $\Omega_R = \mu_{\perp} B_\mathrm{RF}/(2\hbar)$. Here, $\mu_{\perp}$ is the matrix element of the magnetic dipole moment (operator) $\mu_y=g \mu_B S_y$, i.e.~the transition dipole moment that sets the coupling strength between the two spin states~\cite{Perrin2017,SteckBook}${}^,$\footnote{ Note that in general both the transition dipole moment $\mu_{\perp}$ and the differential Zeeman shift $\delta\mu_\parallel$ of the two spin states depend on the magnetic dipole operator and on the quantization field amplitude $B_0$. However, while $\mu_{\perp}$ is the off-diagonal matrix element of the magnetic dipole operator evaluated between the two atomic states, $\delta\mu_\parallel$ is the difference between the expectation values of the longitudinal component of the magnetic dipole $\mu_z=g\mu_BS_z$, parallel to the quantization field $\mathbf{B}_0$.}.
The coupling strength of the RF field with an atomic spin is therefore proportional to the magnetic moment component orthogonal to the quantization axis, which in turn depends on the atomic species and the amplitude of the quantization field $\mathbf{B}_0$, and to the amplitude $B_\perp$ of the RF field component orthogonal to the quantization $z$-axis. The main objective of our circuit design is to maximize the latter quantity for a given set of spatial constraints, while minimizing unwanted interference with the surrounding electronic instruments.

\vspace*{-10pt}
\subsection{Design constraints}
\vspace*{-4pt}

Atomic physics experiments require a RF B-field having the largest feasible amplitude within the sub-mm region of space occupied by the atomic sample. In many setups, atoms are however shielded inside a metallic vacuum chamber, which has only a few optical apertures. A pair of reentrant glass viewports located at cm-distance from the atoms is often included, to grant large optical access for laser beams and high-resolution imaging optics, representing also the best location where to place large-field electromagnets and a RF field applicator. However, in this configuration, the RF coil becomes coaxial with the Feshbach electromagnets, challenging the requirement that the RF field be orthogonal to their magnetic field. 
The RF frequency range of atomic hyperfine transitions, e.g.~$\nu_0\approx 80$\,MHz for the ground-state transitions of lithium atoms, combined with the typical dimensions of the viewports %
and their distance from the sample, determine working conditions characteristic of magneto-quasistatic fields~\cite{HausBook}, i.e. slowly oscillating magnetic fields that behave essentially as static. In our apparatus, the reentrant viewports have a diameter of about 60\,mm and their outer glass surfaces are situated at a distance of 20\,mm from the sample -- to be compared with RF wavelengths of at least few meters.
The standard emitter to generate such quasistatic field at the near-field location of the atomic sample consists of a small coil of conductive wire. 
The amplitude $B_\mathrm{RF}$ in the near-field is proportional to the current flowing in such inductive loop \cite{HausBook}, which must be placed as close as possible to the sample, considering that the field strength of a magnetic dipole decays with the cubed distance.

A possible strategy to minimize the distance between the RF coil and the sample, and to optimally orient the generated B-field polarization $\mathbf{B}_\mathrm{RF} \perp \mathbf{B}_0$, is to accomodate the coil inside the vacuum chamber. This solution has recently become somewhat more common, but it must be planned ahead of constructing the apparatus. Further, it presents issues with flexibility and 
tunability, as well as risks of feedthrough and vacuum pressure failures, %
due to difficult heat management via the vacuum electrical feedthroughs.  
Therefore, we rather consider a wire loop with a diameter of few centimeters placed just outside the chamber within one of the reentrant viewports. At the center of the chamber, this will generate only a B-field, the E-field being negligible
\footnote{In such near-field quasistatic condition, it is indeed inappropriate to refer to the emitter as an ``antenna'', since the propagating electromagnetic radiation is irrelevant, and we will henceforth simply refer to it as the ``coil''.} 
(and unnecessary).
Essentially, the coil transforms the current flowing in it into a B-field, hence in the ideal case no energy is dissipated, the only necessary entity being the current itself. %
Practically, some power is dissipated in the coil itself, in the conductive parts of the vacuum chamber surrounding the coil, and in all resistive elements within the circuit placed between the generator and the coil (e.g.~parasitic resistances associated with transmission line losses). 

\vspace{-7pt}
\subsection{Circuit concept and design}
\vspace{-3pt}

A common approach for maximizing the B-field is to maximize the power transfer from the RF generator towards the coil via impedance matching, by interposing a suitable network of elements between the generator and the coil \cite{WenzThesis,MitraPhD,GeigerPhD}. However, a coil with such small dimensions has only a tiny radiation resistance of (at most) a few tens of m$\Omega$ for VHF frequencies, much smaller than its own loss resistance on the order of 1$\,\Omega$, and negligible with respect to the 50$\,\Omega$ output impedance of the generator. Matching the impedance of the generator is thus particularly challenging, since widely different loads entail very high currents and voltages across the LC elements of the interposed matching network, leading to power losses \cite{CollinBook}. Additionally, since the matching network must be placed at the location of the coil in order to avoid that the feedline becomes part of field-emitter itself, it is highly impractical to adjust the resonant frequency of the circuit, since this location is typically not accessible after the installation in the experimental setup. 

Rather than attempting to directly match the coil impedance to the 50$\,\Omega$ output impedance of the generator\cite{Barker2020}, 
here we adopt a different strategy that up-scales the current flowing through the coil, while avoiding high voltages across matching elements and minimizing radiation from the feedline. The essential idea is to integrate the coil into a transmission line with a low controlled impedance using an impedance transformer, to achieve a high current level in the coil while avoiding large power dissipation through its parasitic losses. Importantly, this approach avoids introducing any unbalanced line segments that would emit unwanted RF fields. This is essential for stable operation and minimizing electromagnetic interference with the experimental instrumentation.

To realize this scheme, we design a setup consisting of a series of elements that work together to form an optimized impedance transformation system. The RF generator, which has an output impedance of 50$\,\Omega$, is connected to a step-down impedance transformer that significantly reduces the impedance seen by the coil. 
Immediately after the step-down transformation, a matching network is introduced to interface the impedance of the coil with the rest of the circuit through a tuning board.
This board also ensures that the system operates at the desired resonance frequency, enabling fine-tuning of the frequency response without requiring physical modifications in the vicinity of the coil. 
Following the coil, a step-up impedance transformer is used to restore the impedance back to 50$\,\Omega$, allowing the majority of the power to be dissipated in a dummy load, rather than mainly in the coil as it would happen by connecting the latter directly to ground. While the 50$\,\Omega$ resistive load decreases the power transferred to the coil and thus the current by a factor of almost 2, it is fundamental to balance the network design, avoiding common-mode currents on the ground return path that would cause unwanted radiation from the cable connecting the RF generator to the step-down transformer. Indeed, with an unbalanced network the transmission-line section located upstream of the first step-down transformer, between the RF generator and the step-down transformer itself, would feature different currents flowing on the inner conductors and the (grounded) shield and would thus radiate efficiently, having a length of tens of cm, comparable with radio wavelengths in the VHF range.
The main blocks of our circuit design are illustrated in Fig.~\ref{fig:schblkconc}(a).
%

\begin{figure*}[!t]
  \begin{center}
    \includegraphics[width=0.75\textwidth, angle=0]{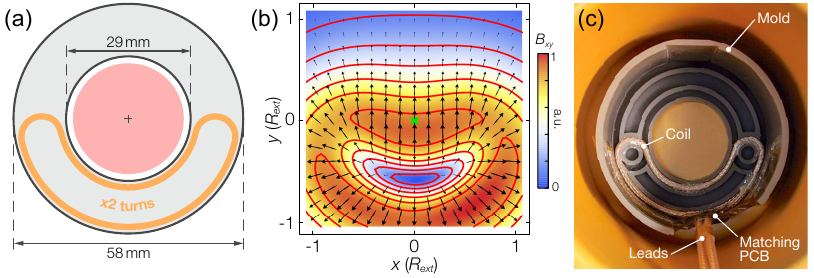}
    \caption{\label{fig:coil} Coil design and realization. (a) The coil is formed by \fs{two turns of} a half-annulus shaped coaxial wire loop \fs{(orange line)}, with an external radius $R_\mathrm{ext} \simeq 29$\,mm. A clear \fs{optical} aperture of 29\,mm diameter, centered on the viewport remains available, \fs{used e.g.~to shine a 25.4\,mm-diameter beam on the atoms (pink circle)}. (b) Calculated B-field polarization pattern and transverse component amplitude $B_{\perp}$ in the atomic plane at $z \simeq 0.7\,R_\mathrm{ext} \simeq 20\,$mm. The in-plane component $B_{\perp}$ is dominant and quite homogeneous around the sample location $x,y=0$ (green cross). 
    In particular, at the largest contour line, that delimits a region of about $30 \times 10 \,$mm$^2$, $B_\perp$ is reduced by only 15\% with respect to its value at the atoms' position.
    (c) The coil has been realized using coaxial RG316 cable, and is held in shape by a 3D-printed supporting mold \fs{(gray plastic cylinder with grooves)}. \fs{The two lead cables visible in the bottom part of the picture connect to the coil through two capacitors C2 and C3 soldered to a cm-sized PCB [see Fig.~\ref{fig:Realiz}(a)].} Measurements of the coil impedance are performed for simulation purposes, surrounding it with a metallic cylinder to emulate the EM environment within the re-entrant viewport.\vspace*{-15pt}}
  \end{center}
\end{figure*}

%
Several approaches exist to transforming a transmission line impedance, including the use of ferrite-loaded binocular transformers and stub-loaded line transformers \cite{CollinBook}. We adopt a quarter-wave coaxial line transformer due to its effectiveness in achieving a high impedance transformation ratio while maintaining simplicity and robustness. In addition, the quarter-wave transformer is particularly suitable for our application because it does not introduce ferromagnetic materials, which would interfere with the high static magnetic fields used in ultracold-atom experiments for interaction tuning, creating uncontrolled magnetic disturbances to the atomic levels. The quarter-wave transformer is a narrowband solution, but it is fully compatible with a typical tunability range of a few MHz required to address the atomic resonance at $\nu_0=\omega_0/2\pi$ for different values of the external magnetic field $B_0$ -- e.g. $\nu_0=81-85$ MHz for the $\ket{2} \leftrightarrow \ket{3}$ transition of ${}^6$Li. 
Moreover, the proposed approach can yield a large amplification of the current flowing through the coil, depending on the choice of the quarter-wave transformer ratio. This solution also guarantees the reproducibility of the performance of the antenna due to its mechanical and electrical stability, reducing the need for later adjustments. 

Summarizing the main benefits of the proposed design: 
(i) it provides a significant enhancement of the current flowing through the coil for a given available power from the generator, ensuring access to the strong RF driving regime of atomic transitions; 
(ii) it allows for straightforward frequency adjustments through a single accessible low-voltage component, facilitating experimental fine-tuning in situ; 
(iii) it minimizes RF interference and ensure compliance with electromagnetic compatibility (EMC) requirements, reducing the risk of unintended radiation affecting other laboratory equipment; 
(iv) it simplifies the overall circuit construction, making it easily adaptable to similar experimental setups.

\subsection{The coil}\label{sec:coil}
A high B-field can be obtained with a coil consisting of several turns, but the self-inductance rises with the square of the number of turns. High values of the self-inductance yield a reactance that is difficult to compensate. In addition, as described previously, only the generated field component oriented orthogonally to the atom quantization axis $z$ couples to the atomic spin, i.e.~$\mathbf{B}_\mathrm{\perp} \perp \mathbf{B}_0$. Since $\mathbf{B}_0$ is produced by strong electromagnets also installed in the reentrant viewports, $z$ is perpendicular to the viewport surface.
We have considered a variety of three-dimensional asymmetric coil geometries, ranging from tilted ellipses to off-centered asymmetric or twisted profiles. We have assessed their performance in simulations by quantifying: (i) the magnitude of the B-field component $B_\mathrm{\perp}$ to maximize the Rabi frequency; (ii) the homogeneity of such transverse $B_\mathrm{\perp}$ field around the location of the atomic sample to minimize spatial dephasing over extended atomic samples; (iii) the clearance along the $z$-axis at $x,y = 0$ to maximize optical access to the atomic sample. 

The best trade-off was found with a two-turn loop shaped as a half-annulus, as depicted in Fig.~\ref{fig:coil}(a), reminiscent of previously reported kidney-shaped loops\cite{LompePhD, MitraPhD}.
This planar coil geometry maximizes the field component orthogonal to $\mathbf{e}_z$ [see Fig.~\ref{fig:coil}(b)] for the fixed minimum distance $d$ between the atoms location and the coil, $|d| \simeq 0.7\,R_\mathrm{ext}$. 
In particular, at the atoms location, the simulation estimates a magnetic-field component $B_{\perp} \approx 0.05\,\text{G/A}_\mathrm{pp}$ and an angle of approximately 70$^\circ$ between the field and the z-axis (see Appendix). This angle becomes 90$^\circ$ at a distance of about half of the coil external radius, which is however incompatible with our coil external radius and distance-to-atoms constraints.
Moreover, the generated field is nearly homogeneous over several millimeters around the atomic sample location [see Fig.~\ref{fig:coil}(b)], rendering the design robust to manufacturing imperfections.
The coil has been realized using the the same coaxial cable used for transmission-line transformers (see next paragraph). Using coaxial cable is not essential, as RF currents flow only on the braid and the central conductor is inactive, 
but it simplifies the construction and makes it reliable even for high powers.
%

\subsection{Circuit implementation and simulation}

\begin{figure}[!b]
  \begin{center}
    \includegraphics[width=0.95\columnwidth, angle=0]{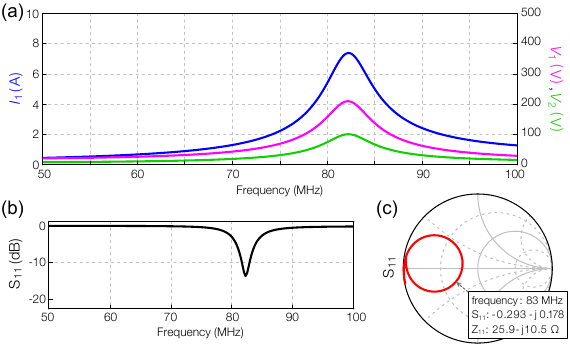}
    \caption{\label{fig:simbanant1} Circuit simulation results for a 100\,W generator, calculated based on the components displayed in Fig.~\ref{fig:schblkconc}(b). (a) The current (peak-to-peak) $I_1$ in the coil (blue), and the voltages (peak-to-peak) $V_1$ (magenta) and $V_2$ (green) across the matching capacitors C1 and C2, respectively, are displayed. (b)-(c) The modulus and Smith chart of the circuit $S_{11}$ parameter are shown, quantifying the power reflected back to the generator.}
  \end{center}
\end{figure}

The circuit is implemented according to the schematic in Fig.~\ref{fig:schblkconc}(b).
Figure \ref{fig:simbanant1} shows a simulation of the whole circuit carried out with the Qucs software\footnote{Available under GPL license at \href{http://qucs.sourceforge.net/}{http://qucs.sourceforge.net}}, which guided the choice of the actual components and parameters. 
To match the impedance from 50$\,\Omega$ (generator) to a lower value suitable to increase the current in the coil, a $\lambda$/4 transmission-line transformer\cite{CollinBook} has been used. 
The relationship between the characteristic impedance of the line $Z_0$, and the impedances seen at the input side $Z_\mathrm{in}$ and output side $Z_\mathrm{out}$ of the transformer is given by:
\begin{equation}
  Z_\mathrm{out} = \frac{Z_0^2}{Z_\mathrm{in}}
\end{equation}
Thus, to have a large impedance variation towards lower values, it is necessary to have low-impedance transmission lines. Lines with characteristic impedance between 10$\,\Omega$ and 75$\,\Omega$ are available, but for simplicity we have chosen to use only 25$\,\Omega$ coaxial cables.
Using two 25$\,\Omega$ coaxial cables in parallel (or equivalently four 50$\,\Omega$ cables), has approximately the same effect of using a single 12.5$\,\Omega$ cable, with the added advantage of power sharing between different cables, such that small ones can be used.
With this choice, the impedance seen after the transformer towards generator is 3.125$\,\Omega$, namely 16 times lower than the original value. 
More dramatic impedance reductions are feasible by adding further cables in parallel, until parasitic resistances of a few 100\,m$\Omega$ of the coil (skin effect) and solder connections may become a limiting factor.


\begin{figure}[!t]
  \begin{center}
    \includegraphics[width=1.0\columnwidth, angle=0]{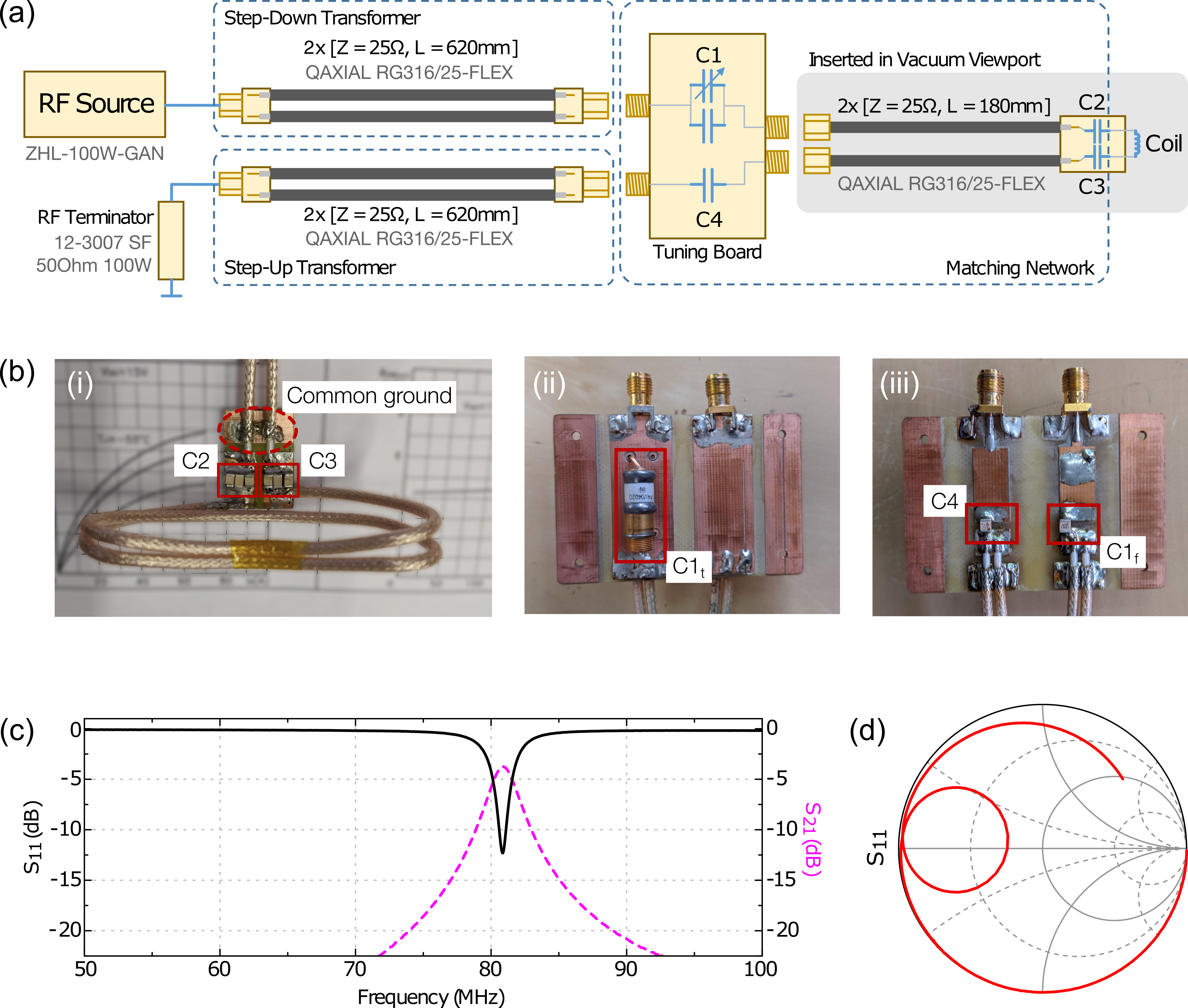}
    \caption{\label{fig:Realiz} Circuit realization and \textit{in situ} measurements. (a) Sketch of the circuit boards and components. (b) Pictures of the realized matching (and tuning) network board. Capacitors C2 and C3 are hosted by a cm-sized PCB mounted vertically next to the coil within the plastic housing [see Fig.~\ref{fig:coil}(c)]. The two coaxial cables connecting the tuning board with C2 and C3 share a common ground, preventing B-fields from being emitted at any other location on the feedline than the loop itself. Capacitors C1 and C4 are soldered on a dedicated PCB, which is housed and fixed to a breadboard outside the reentrant viewport. The tuning capacitance is realized by two capacitors C1$_{\mbox{f}}$ (fixed) and C1$_{\mbox{t}}$ (trimmer), C1 = C1$_{\mbox{f}}\,$+ C1$_{\mbox{t}}$. All fixed capacitors are ATC-100B series, while the tuning capacitance in C1 is realized with a Johanson MAV05D30. (c)-(d) The $S_{11}$ parameter measured with a vector network analyzer is found in good agreement with simulations [see Fig.~\ref{fig:simbanant1}(c)]. The $S_{21}$ parameter quantifies the power reaching the final 50\,$\Omega$ load, and equals about $-3\,$dB at the resonance (here $\simeq~$81\,MHz).
    }
  \end{center}
\end{figure}

The matching network, which includes series capacitors and 25$\,\Omega$ \fs{coaxial} cables and incorporates the coil, serves to cancel the inductive reactance of the coil itself. The 25$\,\Omega$ cable sections also facilitate the installation and the manual capacitance tuning in the intricate environment close to the experimental chamber.
After the last (tunable) capacitor C1, another $\lambda$/4 transformer rises again the impedance from 3.125$\,\Omega$ to
50$\,\Omega$, matching that of the terminating load.
As can be seen from simulations shown in Fig.~\ref{fig:simbanant1}(a), in addition to a good matching over a relatively wide frequency bandwidth (FWHM) of about 10\,MHz, 
the current $I_1$ is enhanced by a factor \smash{$\eta \simeq (Z_\mathrm{in}/\tilde{Z}_\mathrm{out})^{1/2} \lesssim 4$} with respect to the current that would be obtained using a generator transferring the same power onto a generic $50\,\Omega$ load (e.g., 2\,A$_\text{pp}$ for a 100\,W generator). 
Here, $\tilde{Z}_\mathrm{out} \gtrsim 3.125\,\Omega$ denotes the impedance seen at the output side of the step-down transformer taking also into account the 180\,mm-long transmission-line section.

The other graphs show the voltages $V_1$ and $V_2$ across capacitors C1 and C2, which are well below the maximum working voltage of ceramic capacitors such as the ATC 100B series and air dielectric trimmers.
With respect to the well-established approach of L-matching the coil impedance creating a resonant LC network, our design improves the bandwidth of more than a factor of 2 while considerably reducing the voltage across the matching capacitor, at the cost of a lower Q-factor (by a factor less than 2).
In our simulations, the coil has been modeled as a LR series, whose values are based on vector impedance measurements of the actual coil build [see Fig.~\ref{fig:coil}(c)]. Given the low RF frequencies involved and the small dimensions, no parasitic resonance was observed, so the model turned out to be particularly simple -- no parallel capacitance is included. As described in the following paragraphs, measurements and simulations are in close agreement.

\vspace*{-10pt}
\subsection{Realization}

Figure~\ref{fig:Realiz}(a) shows a sketch of the realized circuit. We opted to add an extra transmission-line section of 180\,mm length, to reach the vacuum viewport of the experiment while keeping the tuning capacitor C1 at reach.
The system is built using a $Z_0=25\,\Omega$ coaxial cable (QAXIAL RG316/25-FLEX) with non-magnetic conductors and high-power handling of $\sim$\,500\,W at 80\,MHz, and high-voltage porcelain multilayer SMD capacitors from American Technical Ceramics (ATC 100B series, non-magnetic). All ATC capacitors are able to withstand 1\,kV and have a Q-factor exceeding 1,000 at 100\,MHz.
To hold the coil in shape, a rigid plastic support has been realized with a 3D printer [see Fig.~\ref{fig:coil}(c)], increasing the robustness of the system and
facilitating the insertion of the coil in the vacuum viewport mechanic clearance. The step-down and step-up transformers are realized using two couples of 620\,mm-long coaxial sections, corresponding to $\lambda$/4 segments at 80\,MHz.
The matching network includes [see Fig.~\ref{fig:Realiz}(b)]:
(i) Capacitors C2 and C3, used to create a matching between the loop and the line; (ii) A pair of coaxial cables with characteristic impedance $Z_0=25\,\Omega$ which link the C2/C3 board to the tuning board; (iii) Capacitors C1 and C4, realized with 2 fixed capacitors and a tunable trimmer capacitor, necessary for finely tuning the resonance frequency. 
The latter is implemented by an air dielectric trimmer capacitor from Johanson Technology, with a voltage rating of 500\,V, a capacitance range of 1\,pF -- 30\,pF and a Q-factor $>\,$800 at 100\,MHz.  


\section{Circuit performance}

\subsection{Electrical network measurements}

The measurement phase started with the characterization of the coil impedance to determine the actual L and R parameters to input in the circuit model.
The matching network has been designed, realized and tested on the workbench.
Subsequently, the coil has been inserted in the viewport and the complete system tested \textit{in situ}. Illustrative measurements performed with a vector network analyzer (VNA) are shown in Fig.~\ref{fig:Realiz}(c)-(d). 
The circuit resonant frequency can be shifted over a range of about 2\,MHz varying the value of C4 by 30\,pF without visibly reducing the impedance matching. For a given resonance position, the current bandwidth, defined as the full width at half maximum (FWHM) of the current resonance, equals approximately 10 MHz [see Fig.~\ref{fig:simbanant1}(a)].

\subsection{Testing the system on atomic samples}\label{sec:rabis}
%

\begin{figure}[!t]
\centering
{\vspace*{0pt} \includegraphics[width=0.8\columnwidth]{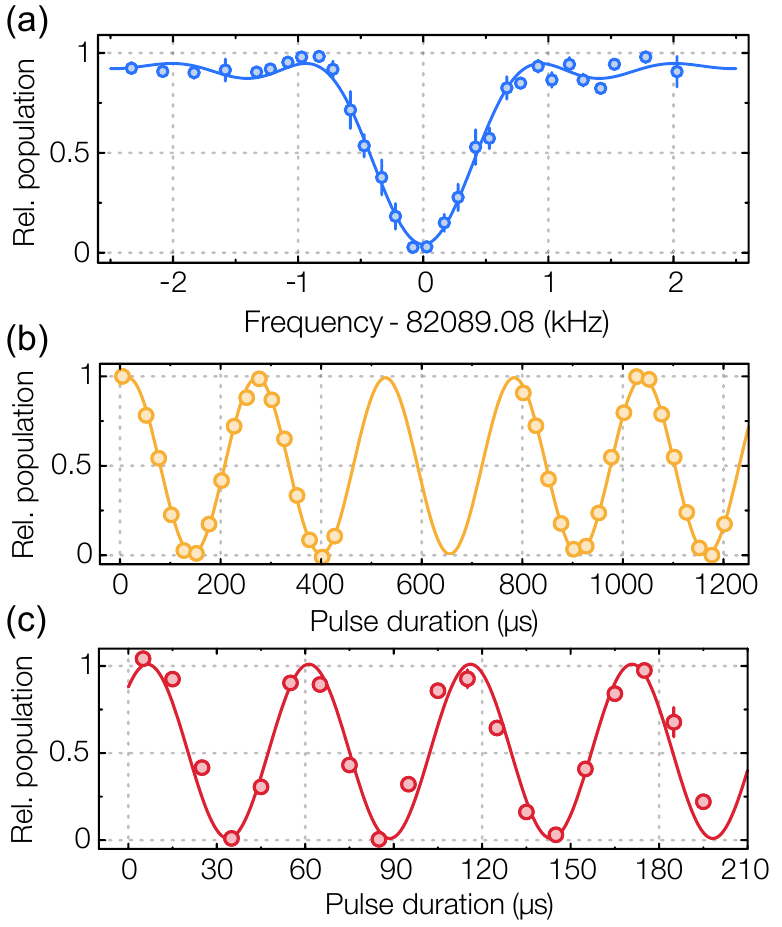}}
	\caption{System performance measured with an ideal Fermi gas of $^6$Li atoms at $B_0 \simeq 770$\,G. On the vertical axis, the relative population denotes the number of atoms in state $\ket{2}$ normalized to the total atom number. (a) Spectroscopy of the $\ket{2} \leftrightarrow \ket{3}$ transition. The solid line is a fit of the data with the expected sinc-function lineshape, yielding a resonance frequency $\nu_0 = 82.08908(1)$\,MHz and a Rabi frequency $\Omega_R/2\pi = 410(16)$\,Hz. (b)-(c) Rabi oscillations at the resonance between states $\ket{2}$ and $\ket{3}$ for RF generator powers of about 3.5\,W (yellow) and 80\,W (red). Solid lines are sinusoidal fits to the data, yielding Rabi frequencies $\Omega_R/2\pi = 3.909(4)$\,kHz and $\Omega_R/2\pi = 18.25(23)$\,kHz, respectively.\vspace*{13pt}}
\label{fig:spectro}
\end{figure}

We have tested the system in our experiment with ultracold degenerate gases of fermionic lithium atoms\cite{Ketterle2008,Amico2018}. We have addressed the transition between the second-lowest and third-lowest magnetic sublevels of ${}^6$Li ground state, usually labeled $\ket{2}$ and $\ket{3}$, respectively.
In the experiment, offset fields in the range $B_0 \simeq $ 570\,G\,--\,900\,G are used \cite{Ketterle2008}, covering the position of two broad Feshbach resonances at 690\,G and 832\,G between states $\ket{2}$ and $\ket{3}$, and the lowest hyperfine level $\ket{1}$.  
For such $B_0$ values, nuclear and electron spins of the atom are largely decoupled \cite{SteckBook,Gehm2003}, and the spin system formed by states $\ket{2}$ and $\ket{3}$ has a small magnetic moment $|\mu_\perp|/h\lesssim\,$70\,kHz/G (still much larger than the nuclear magnetic moment $\mu_N \simeq \mu_B/1836$, with $ \mu_B/h \simeq 1.4\,$MHz/G).  

A spectroscopic measurement of the $\ket{2} \leftrightarrow \ket{3}$ transition in an ideal (spin-polarized) degenerate Fermi gas at $B_0 \simeq\,$770\,G is displayed in Fig.~\ref{fig:spectro}(a). A sinc-shaped line centered around 82\,MHz is observed, associated with the 1\,ms-long square pulse at 50\,mW of RF power, used for the spectroscopic transfer shown here. 
A spectral stability below 100\,Hz is observed during several minutes necessary for a full spectroscopic measurement, corresponding to an offset B-field stability below 10\,mG, or a fractional stability around $10^{-5}$. 

In order to benchmark the performance of our RF system at high power, we have performed Rabi oscillation measurements at the measured resonance of the $\ket{2} \leftrightarrow \ket{3}$ transition. Illustrative results are displayed in Fig.~\ref{fig:spectro}(b)-(c), where a coherent sinusoidal evolution of the state populations is visible for two different driving powers. The intrinsic lifetime of the hyperfine states is much longer than the typical measurement time scales, guaranteeing that visible damping originates only from dephasing from temporal magnetic field fluctuations or spatial inhomogeneities over the atomic sample. Since no measurable damping is observed during the evolution, we conclude that our RF setup does not significantly influence the surrounding electronic equipment, used to actively stabilize the value of $B_0$ (comprised of current transducers, analog PID loops, and DC regulated power supplies) over the typical timescale of the experiment for each RF power. Extracting the Rabi frequency with a sinusoidal fit to the data, we find $\Omega_R/2\pi \simeq 18.25\,$kHz for a driving power of about 80\,W. This corresponds to a $\pi$-pulse duration of $27.5\,\mu$s, which is on the order of the Fermi time $\tau_F \simeq 30\,\mu$s in our degenerate Fermi gas. 
The measured Rabi frequency yields a coupling B-field of 0.29 G, in good agreement with the simulation predictions of 0.32 G.
%

\section{\label{sec:conclusions} Conclusions}
\vspace{-3pt}

In this paper we have described a practical approach to the generation of strong, quasistatic $\sim100\,$MHz RF magnetic fields for atomic physics experiments. This provides an efficient, low-cost and uncomplicated solution to achieve large coupling strengths with atomic spins, without introducing detrimental disturbances or reliability issues. The realized circuit is able to withstand short pulses with RF power $P>100$\,W, further increasing the maximum attainable B-field amplitudes and atomic Rabi frequencies. Alternatively, larger current enhancement factors $\eta$ could be obtained by implementing higher impedance-transformer ratios $\eta^2$, e.g.~$\eta=6$ would allow to achieve the same maximum Rabi frequency demonstrated here by using a $\sim 40\,$W generator.

Our design can be straightforwardly adapted to other alkali atomic species such as K, Rb or Na. Recently, it has been extended to address the RF transition at $\sim 240\,$MHz in Cr atoms \cite{CoscoThesis}, demonstrating the feasibility of our approach even at frequencies beyond the magneto-quasistatic regime\cite{Kraus1988}$\!{}^{,}$\footnote{For frequencies exceeding 200\,MHz, the RF coil features a typical radiation resistance around 1\,$\Omega$, and its behavior crosses over to that of a small-loop antenna \cite{Kraus1988}, for which the radiated power becomes significant.}. 
Moreover, our scheme is flexible against optical apertures or positioning constraints, possibly also for in-vacuum RF coils, where tilted or deformed circular loops may be considered.
We believe our system may become a standard solution for RF field coupling to alkali atom spins in 
current and forthcoming experimental setups, where large optical access and modularity have become a must.

\textit{Note added:} After the first release of this manuscript in the arXiv repository, our circuit design and coil profile have been extended in Ref.~\citenum{wei2025} to the low-frequency regime ($\sim 30 \, $MHz). In this adapted version, the field-emitting coil serves also as the inductive element within a low-loss capacitive transformer, which entirely replaces the step-down quarter-wave transformer in our design.

\begin{acknowledgments}
We  thank Andreas Trenkwalder, Michael Jag, Alessio Ciamei, Matteo Zaccanti, Marco De Pas for useful discussions, Caterina Credi for help with 3D printing, and the LENS Quantum Gases group for constant support. This work was supported by the Italian Ministry of University and Research under the PRIN2017 project CEnTraL, and the PNRR MUR project PE0000023-NQSTI. %
The authors acknowledge support from the European Union - NextGenerationEU for the ``Integrated Infrastructure initiative in Photonics and Quantum Sciences'' - I-PHOQS [IR0000016, ID D2B8D520, CUP B53C22001750006]. This publication has also received funding under the Horizon Europe program HORIZON-CL4-2022-QUANTUM-02-SGA via project 101113690 (PASQuanS2.1).
\end{acknowledgments}

\section*{Data Availability Statement}
The data that support the findings of this study are available from the corresponding author upon reasonable request.

\appendix*\section{Simulated B-field profile}

\begin{figure}[h!]
\centering
{\vspace*{0pt} \includegraphics[width=1.0\columnwidth]{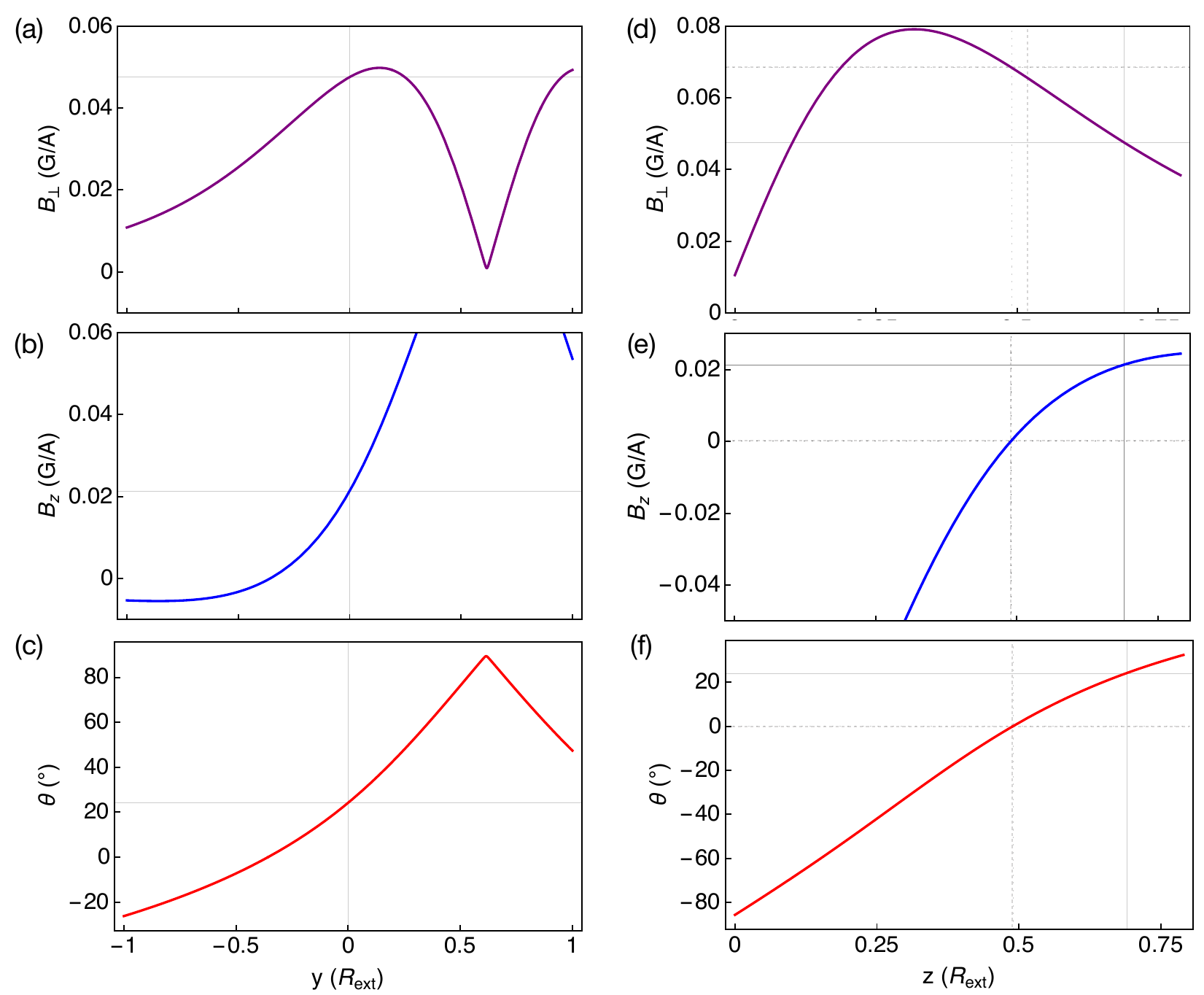}}
	\caption{Simulated spatial profile of the B-field generated by the RF coil for a current of 1\,A$_\text{pp}$ flowing through the coil. (a)-(b) Transverse and axial RF-field amplitudes, $B_\perp$ and $B_z$, along the $y$-axis at $(x,z) = (0, |d|)$. (c) B-field vector angle with respect to the $x-y$ plane along the $y$-axis. (d)-(e) Transverse and axial RF-field amplitudes, $B_\perp$ and $B_z$, along the $z$-axis at $(x,y) = (0, 0)$. (f) B-field vector angle with respect to the $x-y$ plane along the $y$-axis. The position of the atomic sample is $(x,y,z)\simeq (0,0,|d|)$ with $|d|\simeq 0.7 \,R_\text{ext}$, as denoted by solid vertical grid lines. Dashed vertical grid lines denote the optimal coil distance from the atomic sample to obtain a B-field perpendicular to the $z$-axis at $(x,y)=(0,0)$.}
    \vspace*{10pt} 
\label{fig:Bfield}
\end{figure}

In this Section we provide further details on the simulations that we performed to optimize the coil shape. %
Our simulations account for the real shape of the coil (including the finite volume of the wire loops), identifying the half-annulus design as the best trade-off considering the criteria mentioned in Sec.~\ref{sec:coil}. Such a solution offers a clear shape prescription, easy to reproduce and adapt to different absolute dimensions. The field-generating performance remains indeed optimal as long as the target field location lies at $(x,y) \simeq (0, 0)$ and $|z| \in [0.5\,R_\text{ext}, 0.7\,R_\text{ext}]$, see Fig.~\ref{fig:Bfield}. Here, $(x,y,z) = (0, 0, 0)$ refers to the center of the annulus in the mid-plane of the coil itself. 

The simulated transverse $B$-field magnitude at the atoms location, $(x,y,z) \simeq (0, 0, |d|)$, equals $B_{\perp} \approx 0.05 \,$G/A$_\text{pp}$. The axial $B$-field at the same location is estimated as $B_z \leq 0.4 \,B_{\perp}$, corresponding to a field angle with respect to the $x-y$ plane of about $22^\circ$. Decreasing the distance from the atoms to the emitter by around $5 \,$mm would completely zero the axial $B_z$ component, yielding also a moderate increase of the useful field $B_\perp \approx 0.068 \,$G/App. This indicates that an optimal ratio of sample distance to coil annulus external radius is realized for $|d|/R_\text{ext} \approx 0.5$. However, this ratio is incompatible with the constraints of our vacuum chamber, and the smallest achievable ratio in our case is around $0.68$, i.e.~$|d| \approx 20 \,$mm.

%

The simulated values for the transverse $B$-field generated by the coil match reasonably well the values obtained experimentally, which can be extracted from the atomic Rabi frequencies measured shown in Section~\ref{sec:rabis} by calculating the magnetic dipole moment $\mu_{\perp}\approx 70 \,$kHz/G of lithium $2 \leftrightarrow 3$ hyperfine transition. For $80 \,$W and $3.5 \,$W, we measure Rabi frequencies of $18.3 \,$kHz and $3.9 \,$kHz, respectively. These values correspond to coupling $B$-fields of $0.29 \,$G and $0.056 \,$G, respectively, to be compared to simulation predictions of $0.32 \,$G and $0.07 \,$G.

\vspace*{-10pt}
\section*{References}
\bibliography{bobant-bib} 


\end{document}